\documentclass[11pt]{article}
  \usepackage{times} 
  \usepackage{helvet} 
  \usepackage{courier} 
 
  \usepackage{amssymb}
\usepackage{amsthm,graphicx}
\usepackage[noend]{algorithmic}

\newtheorem{lemma}{Lemma}

\newtheorem{theorem}{Theorem}
\newtheorem{observation}{Observation}

\DeclareGraphicsExtensions{.pdf}

  \newcommand{\X}{\mathrm{X}}
\newcommand{\CN}{\mathrm{CN}}
\newcommand{\CG}{\mathrm{CN}}
\newcommand{\SOW}{SOW}

\newcommand{\IMCG}{MAGNET CCC Game}

       \title{Composition Games for Distributed Systems: the EU Grant games \protect \thanks{This work was carried out in and supported by the Technion-Microsoft Electronic-Commerce Research Center.}}
  \author{Shay Kutten \and Ron Lavi \and Amitabh Trehan \\
Faculty of Industrial Engineering and Management\\
  Technion, Haifa, Israel\\
  }

\begin{document}
\maketitle

\begin{abstract}

We analyze ways by which people decompose into groups in distributed systems. We are interested in systems in which an agent can increase its utility by connecting to other agents, but must also pay a cost that increases with the size of the system. The right balance is achieved by the right size group of agents. We formulate and analyze three intuitive and realistic games and show how simple changes in the protocol can drastically improve the price of anarchy of these games. In particular, we identify two important properties for a low price of anarchy: agreement in joining the system, and the possibility of appealing a rejection from a system. We show that the latter property is especially important if there are some pre-existing constraints regarding who may collaborate (or communicate) with whom.

\end{abstract}

\section{Introduction}

It is likely that agents will form groups  when they  gain from cooperation. 
In  \emph{peer-to-peer} (P2P) systems, agents team up to share content; in multi-agent systems, agents team up to complete a task. This motivation to cooperate is a ``force'' that pushes towards grouping, and has been addressed in numerous papers (some are surveyed below). Here, we are interested in the combination of this force with a second force that breaks large groups into smaller ones.  A group that is too large may incur costs that are too high, such as overheads, free loaders, exposure to outside threats (e.g.~lawsuits over intellectual properties), etc. This may decrease the value agents get from a large group and motivate them to break it. Also, consider biological ecosystems: 
A species with too few members will die out  due to competition and under exploitation of resources; if the species has too many members and the resources are fixed or slow to renew, the species may die out due to overexploitation. This suggests that for many systems, the sweet spot lies somewhere in the middle.

Consider the example of FP7, the current, 7th Framework Programme for supporting research in Europe. One of its main goals is to form large ``networks of excellence''. Of course, the commission is not satisfied with size alone, it is interested in the combination of size and quality. How can the commission obtain a ``network'' satisfying both of these criteria? Clearly, the commission cannot simply choose some set of researchers to its liking, telling them that they now form a research network, and must cooperate to obtain good research results. Usually, it is hard for the commission to find a researcher's value unless the researcher exhibits this value first, by submitting a proposal. In addition, the values of researchers materialize when they work with researchers with whom they find common grounds and like to cooperate.

This scenario is similar to other cases where agents in a distributed system decompose into smaller groups. For example, the factors mentioned above seem to be present in a P2P system as well. First, people form such systems voluntarily. Second, there may not be a grant, but the members do realize some benefit from pooling their resources (music, movies, etc.) together. Third, we already noted that a motivation is often present, for these people to form a more exclusive system, rather than to have a very large one.

To capture these and other realistic settings, we study a model in which a designer seeks to choose a connected subset of nodes in an underlying network, where each node has a quality parameter. A group of nodes can perform the task only if the sum of their qualities passes a certain threshold, T (we term such a group ``eligible''). Having passed this threshold, the success of a group positively depends on its average quality. The designer (granting agency) wishes to award a grant of M Euros to the best such group.

The winning group should also decide how to partition the grant money among its members. While it may be possible for the agents to bargain on this issue, in practice many times this is not the case. Probably due to strong social norms, researchers simply split the money evenly. This happens in many other settings as well. In P2P systems, the prize is the shared content, to which everyone has equal access. In this paper, we study group composition only under the assumption that the grant is split evenly. There is no doubt that other ways to split the grant are possible, but we defer the study of this issue to future research.

In this work we introduce three natural protocols/games for deciding the composition of groups, and study their {\em price of anarchy} (POA):~\cite{nisan2007algorithmic} the ratio between (a)  the optimal (maximal) average quality of an eligible set of researchers, and (b) the lowest average quality of a winner set that can be formed in an equilibrium (here we use both the notions of a Nash equilibrium and of a strong equilibrium). 
POA is a measure of the degradation of the efficiency of a system due to selfish behavior of agents;  higher ratio corresponds to higher loss and poorer quality  worst case equilibria.
We remark that the use of non-cooperative game theory fits our goals better, compared to using cooperative game theory.  The latter theory is mainly concerned with the correct and most efficient way to distribute payoffs among members of a winning coalition, hiding the workings of {\em how} the winning coalition is formed. However, our interests are different 
and focus on the question: {\em what is the composition of the winning group, and what is its quality}? Moreover, in our model the answer to ``which group wins'' depends not only on the group, but also on the competing groups.

The first protocol, the gold rush game, is a naive composition method often used in some legacy systems (e.g.~mailing lists): joining a group is done by simply declaring (unilaterally) the will to do so. Granting agencies do not usually use such a method. Perhaps the reason is that, as we show, its price of anarchy is very high, bounded only by the size of the whole society. The analysis of this game is trivial, but forms a basis for comparison, and is a good warmup.

Usually, granting agencies require some stronger condition for researchers to join a group: at the least, that all group members agree on its composition (the list of participants). If this method is used and the underlying \emph{collaboration network} is a clique (everyone knows everyone else), the \emph{strong price of anarchy}  improves drastically, to be at most 2. While this improvement is impressive, we also show that when the underlying collaboration network is arbitrary (in particular, {\em not} a clique), the strong price of anarchy of this method can grow up to 3, which means that only a third of the optimal average quality may be realized.

While the source of the high price of anarchy of the first method was the ``too easy'' joining, the source of the (still rather) high price under the second method is just the opposite -- the difficulty of joining. That is, we show examples where some winning set finds it beneficial to disallow the joining of some high value researchers. We introduce a third protocol as a simple alleviation of the previous problems: allowing those rejected high value researchers to ``appeal'' the rejection. (Interestingly, a granting agency named
MAGNET, of the Israeli ministry of industry, uses a similar method for the joining of companies to a consortium). We show that the strong price of anarchy of this third method is lower: at most 2 on arbitrary networks, and even approaches 1 (the optimum) for large sets, if the collaboration network is a complete graph. The analysis of this third method constitutes the main technical contribution of this paper. It shows how the topology of the collaboration network affects quality, and what is the importance of an appeals process.



\section{Related Literature}
Probably, the most related paper to ours is~\cite{pod}. They, too, use non-cooperative game theory. Moreover, in their model, too, agents have values, and the sum of the values of agents in a winning coalition must exceed a fixed threshold. However, that model seems to focus on points that do not capture our motivating scenarios. First, we, intentionally, focus on the competition part, and assume that the payoff to the winning group is fixed. In contrast, there, the values of the players in the winning coalition are also the total payoff of that coalition. This assumption drastically affects the nature of the competition. Second, we, intentionally, focus on situations where the social norm or physical reality dictates the equal sharing of the grant money; our aim is to characterize the coalitions (and qualities) that will result from the given division of the total gain. In contrast, they aim to analyze the bargaining process through which the division of the total gain from the cooperation will be determined. Finally, our price of anarchy evaluates the values of the winning group. In contrast, the price of democracy they analyze becomes meaningless in our case where there are no costs (costs are an orthogonal parameter introduced there).

\cite{M77} studies cooperative games over graphs, where only connected coalitions $S$ are able to extract their value $v(S)$. This is also one of our assumptions. They show that the unique fair way to divide the value of the grand coalition is by the Shapley value. The current paper does not deal with dividing the value. Moreover, here, a group may or may not win, as a function of the the actions of the players in competing groups. This does not seem to be captured well by cooperative game theory.  

While most works in classic cooperative game theory are only remotely related to our specific model, a series of papers on the stability of coalition structures, see for example \cite{D94}, \cite{BJ02}, and references therein, is quite relevant to our work. They study a setting where society splits into different coalitions, and characterizes cases where such partitions are stable. (This seems more related than the famous stable marriage problem~\cite{stable-marriage}.) Unlike the current paper, they assume that {\em all} formed coalitions win
a prize (otherwise, clearly, no stable coalition structure will emerge). Another difference is that  these papers do not focus on specific protocols for  forming coalitions; their main interest is in characterizing when such stable structures exist. \cite{konishi1997pure} models the creation of coalitions in a game that bears similarities to our initial example game (the gold-rush). However, they do not aim to analyze the quality of the resulting coalition.

Our paper is also related to the literature on network creation games, starting with \cite{network-creation-game}. These papers study games in which nodes decide how to form links in order to create a connected network, and the price of anarchy is analyzed under various assumptions. In their models, the society becomes {\em connected}. In contrast, in our paper, the society {\em splits} to give birth to a {\em strict subset}, and the question is whether the quality of the formed group is far from optimal because of various strategic issues. The  literature on price of anarchy is rich, see~\cite{nisan2007algorithmic} for a survey.

\section{Two Protocols with Opposing Policies}

\label{sec: goldrush}

A granting agency wishes to award a prize of $M$ Euros to a subset of a society of $n$ researchers. Each researcher $i$ has a value $v_i$ that represents her overall quality. A subset of researchers is {\em eligible} if (i) their sum of values is at least some given threshold $T$, and (ii) they form a connected component of the underlying Collaboration Network (CN).
For simplicity, in most of this section, we assume that the underlying collaboration network is the complete graph, but remove this assumption in subsequent sections.
 The granting agency aims to award the prize to a set of researchers (``consortium'') with maximal average quality among all eligible consortia. To exclude some trivial cases, we assume throughout that $v_i < T$ for every researcher $i$, i.e.~the researcher with the maximal value cannot
take the prize on her own. We also assume, without loss of generality, that the sum of all values is larger than $T$. While the agency does not know the values of the researchers, we assume that it can verify the values when a set of researchers submits evidence of their value (this is the ``grant proposal''). The agency constructs a protocol, by which researchers form candidate consortia, and the best formed consortium wins and receives the prize. Two natural protocols give some intuition for possible causes of a high price of anarchy.

\medskip

\noindent
{\bf  The Gold-Rush Game. } This protocol, as well as its analysis, are trivial. However, they serve as a basis for comparison, as well as a ``warm up example''. Each researcher submits a separate proposal, reporting (along with a proof of the researcher's value) some label,
the ``consortium name''. The labels are taken from some finite set of labels $L$. Researchers who report the same label are understood to  belong to the same consortium. The agency awards the prize to an eligible consortium with the maximal average value (in case of ties, any arbitrary (possibly randomized) tie-breaking rule can be used). Each researcher in the winning consortium receives an equal share of the prize.

In terms of game theory, the {\em strategy} of each researcher $i$ in this game is the label $\ell_i$ she chooses. The {\em utility} $u_i(\ell_1,...,\ell_n)$ of  $i$ is 0 if ``her'' consortium loses, and $\frac{M}{y}$ if her consortium wins, where $y$ is the size of the winning consortium. A tuple of strategies $\ell_1,...,\ell_n$ is a {\em Nash equilibrium} if $u_i(\ell_1,...,\ell_n) \geq u_i(\ell_1,..., \ell_{i-1}, \ell'_i,
\ell_{i+1},...,\ell_n)$ for every $i=1,...,n$ and every $\ell'_i \in L$. In other words, in a Nash equilibrium $\ell_1,...,\ell_n$, the utility of each researcher $i$ is maximized by declaring $\ell_i$, given that the other researchers declare $\ell_{-i} = \ell_1,..., \ell_{i-1}, \ell_{i+1},...,\ell_n$. It has become standard in the algorithmic game theory literature to measure the quality of a game/protocol by its {\em price of anarchy (POA)} \cite{price1}. In our case, this is the optimal (largest possible) average value divided by the average value of the winning consortium in the worst Nash equilibrium. (This reflects a worst-case point of view).

Unfortunately, the price of anarchy of the gold-rush game is very high. To show this, it suffices to study the case of distinct values (i.e.~no two values are equal) and a complete CN. To analyze the price of anarchy, the next lemma characterizes all Nash equilibria of this game.
(Most proofs in this section are deferred to the journal version of this paper (arxiv version at~\cite{KuttenLaviTrehanEUGamesArxiv})). 

\begin{lemma}
\label{lem:unique}
Assume that values are distinct and the CN is the complete graph. Then, in every Nash equilibrium
of the gold-rush game either no consortium forms, or all researchers declare the same label, hence
all researchers win.
\end{lemma}

\noindent
This immediately implies an unbounded price of anarchy:

\begin{theorem}
The price of anarchy of the gold-rush game is (arbitrarily close to) $n/2$.
\end{theorem}

There is another, more conceptual problem with the gold-rush game. In reality, researchers (as well as P2P users) may know each other, and can coordinate a joint deviation from the presumed equilibrium strategy, e.g.~the top-value researchers may coordinate to belong to an exclusive consortium. The notion of a Nash equilibrium does not allow such coordinated deviations, and is therefore conceptually weak for our case. A better notion is a {\em strong (Nash) equilibrium}, which requires that no subset of the players can jointly deviate and increase each of their utilities~\cite{strong-eq}. Formally, a tuple of strategies $\ell_1,...,\ell_n$ is a {\em strong equilibrium} if for any $\ell'_1,...,\ell'_n \in L$ there exists a player $i$ such that $\ell'_i \neq \ell_i$ and $u_i(\ell_1,...,\ell_n) \geq u_i(\ell'_1,...,\ell'_n)$.
 However, consider the situation where there exists  at least one eligible group that is a strict subset of the society. One of these groups has maximal average value and should be a winner. Informally, all researchers not in this group will  like to switch their labels to the winners but the winners will like to defect as a group to a different label by themselves. This leads to the following lemma:

\begin{lemma}
If there exists an eligible group which is a strict subset of the society, there does not exist even a single strong equilibrium in the gold-rush game.
\end{lemma}

\noindent
{\bf Consensual consortium composition (CCC). } Intuitively, the bad price of anarchy of the gold-rush game resulted from the fact that it was ``too easy'' for anybody to join a consortium of her liking. The following Consensual Consortium Composition (CCC) game is a first attempt to fix the problems of the previous naive design. In this game, each player submits a ``proposal:'' her value and a list of the researchers in her consortium. An eligible consortium of researchers X then satisfies (1; consistency) each researcher in X submitted X as her consortium, (2; threshold) $\sum_{i \in X} v_i \geq T$, and (3; connectivity) the consortium is connected in the underlying CN. The winning consortium is an eligible consortium with maximal average value. If several such consortia exist, the winning one has minimal size.

As discussed above, Nash equilibrium is not really appropriate in our context. In fact, for the CCC game, a Nash equilibrium is meaningless.
The reader can verify that in this game, any partition of the players into consortia will constitute a Nash equilibrium. We focus on the stronger
and more appropriate notion of a strong equilibrium. Analogous to PoA, strong price of anarchy (SPOA) \cite{strong-price-of-anarchy} in our case is the ratio of the largest average value to the average value of the winning consortium in the worst strong equilibrium.

\begin{theorem}
\label{thm-spoa-ccc-game}
Assume that CN is a clique. Fix an arbitrary tuple of researcher values, and suppose that a minimal eligible consortium with the highest average value has size $k$. Then, the strong price of anarchy of the CCC game is (arbitrarily close to) $1 + \frac{1}{k-1}$. In particular, the SPOA of the CCC game is at most 2.
\end{theorem}

\begin{proof}
Assume the CN is the complete graph. 
Fix a tuple of values $v_1,...,v_n$. Let $OPT$ be a minimal eligible consortium with highest average value, and denote $|OPT|=k$. Let $i \in OPT$ be a player with a lowest value among all players in $OPT$. Then, $\sum_{j \in OPT \setminus \{i\}} v_j < T$: if all values in $OPT$ are equal, this follows from the minimality of $OPT$. If not all values in $OPT$ are equal the inequality follows since otherwise $OPT \setminus \{i\}$ is an eligible consortium with a higher average value than $OPT$. The average value of $OPT \setminus \{i\}$ is, therefore, at most $\frac{T}{k-1}$. Thus, the average value of $OPT$ is also at most $\frac{T}{k-1}$.

Now, let $W$ be the winner consortium in some strong equilibrium. If $|W|=l<k$ then there exists an eligible consortium with $l<k$ researchers. Thus, the $l$ players with $l$ highest values form an eligible consortium with average value not smaller than that of $OPT$, a contradiction. If $|W|>k$ then $W$ cannot be a strong equilibrium. This is because players in $OPT$ can deviate, form a consortium, win, and increase their utility (since the size of the winning subset strictly decreases). Thus, $|W|=k$. Since the sum of values of players in $W$ is at least $T$, the average value of players in $W$ is at least $\frac{T}{k}$.

By the previous conclusions, the ratio of the average value of players in $OPT$ to the average value of players in $W$ is at most $\frac{k}{k-1}$. This proves that the SPOA of the CCC game is at most $1 + \frac{1}{k-1}$.

To prove a matching lower bound, an example (for every given $k$) suffices. For this purpose, consider the following tuple of values, for any small enough $\epsilon > 0$. There are $k$ researchers $1,...,k$, all with
the same value $\frac{T}{k-1} - \epsilon$, and researcher $k+1$ with value $k \epsilon$. In this case, the eligible consortium with the highest average value (of $\frac{T}{k-1} - \epsilon$) is $\{1,...,k\}$. Having researchers $\{1,...,k-1,k+1\}$ form one consortium, excluding researcher $k$, is a strong equilibrium -- one can verify that no subset can deviate and strictly increase each of their utilities. Thus, the SPOA in this case approaches $\frac{k}{k-1}$ as $\epsilon$ approaches $0$.
\end{proof}

\begin{figure}[h!]
\centering
\includegraphics[scale=0.5]{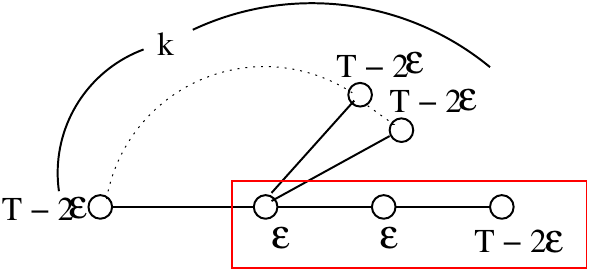}
\caption{ The SPOA for a CCC game on a CN can be arbitrarily close to 3. Here, the nodes are labeled with their values: T is the threshold, $\epsilon$ is an arbitrarily small value. The worst equilibria is shown in the box: the other nodes  and the central $\epsilon$ node form the optimal group.}
\label{fig: CGF-POA}
\end{figure}

When the CN is not a complete graph, this theorem is not necessarily true. Figure~\ref{fig: CGF-POA} shows an example of a non-clique CN
and player values such that the SPOA is $3 - \epsilon$, where $\epsilon$ is an arbitrarily small constant. We conjecture that $3$ is the correct bound. We remark that for the CCC game a strong equilibria always exists, see Lemma~\ref{lm: MagnetSE}. The figure demonstrates another interesting phenomenon: the optimal consortium is not necessarily a strong equilibrium. Here, the central node will prevent the formation of the optimal group.

\section{Main Result: \IMCG\ for arbitrary Collaboration Networks}
\label{sec-imcg}

As shown above, the quality of the winning group of the CCC game may be only one third of the optimal quality. We show how to improve the SPOA to 2 for any collaboration network;  for the complete graph this it will become close to 1. For this purpose, we introduce an extension of the CCC game which proceeds over multiple rounds (the name is inspired by a policy of the MAGNET Israeli granting agency which uses a similar policy). The \IMCG\ is defined as follows:

\begin{itemize}

\item In round 1, execute the CCC game. Let the winning consortium be $W_1$. In each round $r>1$ the winning consortium $W_r$ is an expansion of $W_{r-1}$.

\item In round $r>1$, each researcher not in $W_{r-1}$ can ``submit an appeal'' -- a proposal consisting (as in CCC) of evidence of value and a list of researchers in her consortium. The winning consortium $W_r$ in round $r$ is the union of $W_{r-1}$ and all appealing consortia $\X$ that satisfy
(1; connectivity) $\X \cup W_{r-1}$  is a connected component in $\CN$,
(2; consistency), each researcher in $\X$ submitted $\X$ as her consortium, and
(3; Improvement) $avg(\X \cup W_r)\ >\ avg(W_{r-1})$.

\item The game ends if $W_r = W_{r-1}$ (no justified appeals).

\end{itemize}

\noindent
We next analyze the SPOA of this game, first for any arbitrary $\CN$, then for specific graph structures.

\subsection{Analysis of SPOA for arbitrary $\CN$}

This section shows the main technical result of the paper: The \IMCG\ has SPOA that is equal to exactly 2, regardless of the topology of the CN. To prove this, we first identify some properties that any winning consortium in the \IMCG\ must have. Throughout, we denote by SOW (Social Optimum Winner) a minimal eligible consortium among all eligible consortia with maximal average value.

\begin{lemma}
\label{lm: alltheSEs}
Let $Z$ be the winning consortium in some strong equilibrium outcome of some arbitrary instance of the \IMCG. Then,
\begin{enumerate}
\item \label{lmSE: p1} $Z \cap SOW \neq \emptyset$
\item  \label{lmSE: p2} $|Z| \le |SOW|$
\item \label{lmSE: p3}  $avg(Z) \ge avg(\SOW \setminus Z)$
\end{enumerate}
\end{lemma}

\begin{proof}

\begin{enumerate}

\item $Z \cap SOW \neq \emptyset$:
\label{lmpart: Znonempty}
If Z is not an SOW, $Z$ can win only if it can prevent the formation of $SOW$. This can happen only if $Z$ has some member(s) of $SOW$ i.e.~$Z \cap SOW \neq \emptyset$.

\item  $|Z| \le |SOW|$: If $|Z| > |\SOW|$, the players in  $Z \cap \SOW$ can improve their utility by forming the smaller consortium $\SOW$ in the first round, which is a sure winner (having the highest average). Thus,   $|Z| \le |\SOW|$.

\item  $avg(Z) \ge avg(\SOW \setminus Z)$:
\label{lmpart: Zhighavg}
If $avg(Z) < avg(\SOW \setminus Z)$, the players in $\SOW \setminus Z$ can become winners which is a contradiction. They can appeal together after the currently last round. Since both $SOW$ and $Z$ are connected, so is $\SOW \cup Z$. Since $avg(Z) < avg(\SOW \setminus Z)$, $\SOW \setminus Z$ can be added as winners.

\end{enumerate}

\vspace{-2mm}

\end{proof}

\begin{lemma}
\label{lm: MagnetSE}
The \IMCG\ always has a Strong Equilibrim (S.E.)
\end{lemma}
\begin{proof}
If $\SOW$ is not a S.E., there is a winning consortium $Z$ having non-empty intersection with $\SOW$ and of size strictly smaller than $\SOW$. If $Z$ is not a S.E., some nodes of $Z$ can deviate. I.e., these nodes (maybe with other nodes) can form a consortium $Z'$ having non-empty intersection with $\SOW$ and of size strictly smaller than $Z$. Since this process is finite, we must reach a subset $Z''$ that is a S.E.
\end{proof}

\begin{theorem}
\label{lm: MagSPOA2}
The SPOA of \IMCG\ $\leq 2$.
\end{theorem}

\begin{proof}
Denote $|SOW| = k$. By definition,
\begin{equation}
\label{eqn: poa}
SPOA  =  \frac{avg(SOW)}{avg(Z)}
  = \frac{ \frac{sum(SOW \cap Z)}{k} + \frac{sum(SOW \setminus Z)}{k}}{avg(Z)}
\end{equation}

\vspace{2mm}

\noindent
We prove two properties:

\begin{enumerate}
\item \label{lpf: numr1} $sum(SOW \cap Z)/k \leq avg(Z):$ obviously $sum(SOW \cap Z) \leq sum(Z)$.
By Lemma~\ref{lm: alltheSEs}, $|Z| \le k$. Thus,
\[
\frac{sum(SOW \cap Z)}{k} \leq  \frac{sum(Z)}{k} \leq  \frac{sum(Z)}{|Z|} = avg(Z)
\]
\item \label{lpf: numr2} $sum(SOW \setminus Z)/k \leq avg(Z):$ By Lemma~\ref{lm: alltheSEs},
$avg(Z) \ge avg(SOW \setminus Z)$. Thus,
\begin{eqnarray*}
\frac{sum(SOW \setminus Z )}{k} & \leq &  \frac{sum(SOW \setminus Z )}{| SOW \setminus Z | } \\
& =  & avg (SOW \setminus Z ) \leq  avg(Z)
\end{eqnarray*}
\end{enumerate}

\noindent
Plugging into equation~\ref{eqn: poa}, we get:
\begin{eqnarray*}
SPOA & = & \frac{sum(SOW \cap Z)/k + sum(SOW \setminus Z)/k}{avg(Z)}\\
 &  \leq & \frac{avg(Z) + avg(Z)}{avg(Z)} = 2
\end{eqnarray*}

\end{proof}

\begin{figure}[h!]
\centering
\includegraphics[scale=0.75]{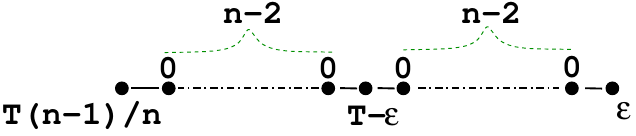}
\caption{For the \IMCG\, a Collaboration Network with a price of anarchy arbitrarily close to 2.}
\label{fig: poaexample}
\end{figure}

\noindent
Figure~\ref{fig: poaexample} shows an example with $SPOA = 2 - \epsilon$, where $\epsilon$ is an arbitrarily
small constant. The $SOW$ consists of the two players with values $\frac{T(n-1)}{n}$ and $T-\epsilon$,
coupled with their intermediary players that have zero values. The worst strong equilibrium here is the
consortium containing the two players with values $\epsilon$ and $T-\epsilon$, coupled with their intermediary
players that have zero values. This gives $SPOA = \frac{2n - 1}{n} - \frac{\epsilon}{T}$, which can be made
arbitrarily close to 2.

\subsection{Dependency on graph parameters}
\label{sub:graph parameters}

The previous analysis showed a SPOA of 2 for arbitrary graph structures. We show how the SPOA may depend on the {\em structure} of the graph. In particular, we consider two extreme special cases: the complete graph on one hand, and the line network on the other hand. These two cases result from differences in the diameter and the connectivity. 

\vspace{0.4cm}

\noindent
{\bf The Complete Graph. } In the CCC game, this case has SPOA $1 + \frac{1}{k-1}$ ($k = |SOW|$). The multiple round version improves this bound to be $1 + \frac{1}{k}$. While the improvement is small for large $SOW$'s, for small $SOW$'s it is quite significant, e.g.~the SPOA can decrease from 2 (with one round) to 1.5 (with multiple rounds).

\begin{theorem}
\label{thm-IMCG-complete-graph}
The SPOA of the \IMCG\ over a complete $CG$ is exactly $1 + \frac{1}{k}$,
where $k=|SOW|$.
\end{theorem}
\begin{proof}
Assume without loss of generality that $v_1 \ge v_2 \ge \cdots \ge v_n$. The proof relies on the following two observations  (full proof deferred to the journal version):

\begin{observation}
For the case of the complete graph, every $SOW$ consortium has size $k$, where $k$ is such that
$\sum_{i=1}^{k-1} v_i < T$ and $\sum_{i=1}^{k} v_i \ge T$.
\end{observation}

\begin{observation}
For the case of the complete graph, the size of any winner consortium in a strong equilibrium outcome
must be equal to the size of the SOW consortium.
\end{observation}

\vspace{-4mm}

\end{proof}

\begin{lemma}
There exists an instance of the \IMCG\ over a complete $\CG$ for which the strong
price of anarchy is arbitrarily close to $1 + \frac{1}{k}$, where $k=|SOW|$.
\end{lemma}
\begin{proof}
Fix $k$, and consider the following tuple of values, for any $\epsilon > 0$: There are $k-2$ researchers $1, \ldots, k-2$ with the same value $\frac{T}{k-1}$. Researchers $k-1$, $k$, $k+1$ have values $\frac{T}{k-1} - \epsilon$, $\frac{T}{k}$, and $\epsilon$ respectively. In this case, the eligible consortium with the highest average is $1,\ldots,k$ having average $\frac{T + T/k - \epsilon}{k}$. The worst strong equilibrium has the winner as the consortium $1, \ldots, k-1, k+1$. This has average  $\frac{T}{k}$, hence the strong price of anarchy approaches $1 + \frac{1}{k}$ as $\epsilon$ approaches 0.
\end{proof}

\vspace{2mm}

\noindent
{\bf The Line Network. } This is the other extreme. Here, the SPOA {\em grows} and approaches 2 as the size of the SOW increases (Theorem~\ref{th: SPOALine}). In contrast, in a complete graph (as shown above), the SPOA {\em shrinks} and approaches 1 as the size of the SOW increases. Intuitively, it seems that this 100\% increase (from the optimum) in the case of a line results from the growth of the diameter when $k$ grows. On the other hand, the disappearance of the price of anarchy (convergence to 1) in the case of a complete graph seems to result from the increase in connectivity.

\begin{theorem}
\label{th: SPOALine}
In the \IMCG\ over a line $\CG$, the SPOA is (arbitrarily close to) $1 + \frac{k-1}{k}$ ($k=|SOW|$).
\end{theorem}
\begin{proof}
Let $W$ be the winner in a worst strong equilibrium, and let $k' = |SOW \setminus W|$. By Lemma~\ref{lm: alltheSEs}, $k' \leq k-1$ (since $SOW \cap W$ is not empty), and $avg(SOW \setminus W) \leq avg(W)$. Hence, $sum(SOW \setminus W) \leq k' \cdot avg(W)$. By the same lemma, $|W| \leq k$, and $k \cdot avg(W) \geq sum(W) > sum(SOW \cap W)$. All these imply:
\begin{eqnarray*}
SPOA  & =  &\frac{avg(SOW)}{avg(W)} \\
 & = & \frac{\frac{sum(SOW \setminus W)}{k} + \frac{sum(SOW \cap W)}{k}}{avg(W)} \\
& \le & \frac{\frac{k' \cdot avg(W)}{k} + \frac{sum(SOW \cap W)}{k}}{avg(W)}\\ 
& = & \frac{k'}{k} + \frac{sum(SOW \cap W)}{k \cdot avg(W)}
\le \frac{k-1}{k} + 1.
\end{eqnarray*}
The other direction is shown via the example in Figure~\ref{fig: poaexample}, using $|SOW| = n$ (specifically, the SPOA there is $\frac{2n - 1}{n} - \frac{\epsilon}{T} = 1 + \frac{k-1}{k} - \frac{\epsilon}{T}$ for any arbitrarily small $\epsilon > 0$).
\end{proof}

\subsection{Additional Issues}
\label{sec: addissues}
We briefly discuss the following two important issues:
\subsubsection{Using a notion of strong subgame perfect equilibrium.}
Since the \IMCG is an extensive-form game, one may wonder whether it is more appropriate to use a notion of Strong Subgame Perfect Equilibria (SSPE) instead of Strong Equilibria. We note that such a change will not change our price of anarchy analysis. First, clearly, every SSPE is also a SE. Thus, using SSPE instead of SE can only decrease the price of anarchy. Second, we note that if $W$ is the winner consortium in some strong equilibrium, then there is also a Strong Subgame Perfect Equilibrium of the \IMCG\ in which $W$ is the winner consortium. This is the tuple of strategies where the researchers in $W$ submit a proposal together already in the first step of the game. In this there will be only one step, and the two notions become the same. Therefore, the price of anarchy is the same, regardless of which notion we use.
\subsubsection{Strong Price of Stability}
\begin{figure}[tbh]
\centering
\includegraphics[scale=0.8]{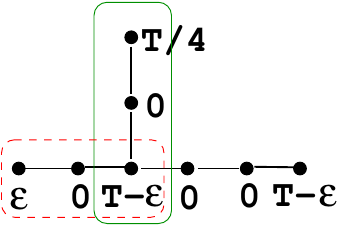}
\caption{The Strong Price of Stability for the \IMCG\ can be greater than 1. The figure shows the best Strong Equilibria (green, solid lines) that yields a SPOS of 6/5. For comparison, the worst Strong Equilibria is shown in red, dashed lines showing a SPOA of 3/2.  }
\label{fig: MagSPOS}
\end{figure}

Another useful concept analogous to the price of anarchy is the price of stability (POS)~\cite{Anshelevich-Price-of-Stability}. Analogous to SPOA, one can define Strong Price of Stability (SPOS) as the ratio of the optimum to a best strong equilibrium. In our game, this is the ratio of the optimal (largest possible) average value to the average value of the winning consortium in the {\em best} strong equilibrium. We do not analyze SPOS in detail in this paper, but we wish to note via an example described in Figure~\ref{fig: MagSPOS} that in \IMCG, the SPOS can be greater than 1.
 
\section{Conclusions and Future Work}

This paper looks at the process of agents teaming up to construct distributed systems. Our setting addresses a specific scenario where one driving force/ incentive limits the size of the consortium,
but another increases it.  We made some simple assumptions. 
We assumed that the value of a researcher is independent of the members of its consortium. We also assumed that the Euro amount of the grant is fixed. What if these assumptions did not hold? What if the grant were some function of the set size?

There are many other interesting directions to explore. We could have more sophisticated utility functions or game designer goals, or we could study ``natural'' games (i.e. not design mechanisms but look at existing systems). We can study more involved environments; such as an evolving dynamic environment where new researchers are born and old ones retire. What about composition of multiple systems? Could multiple consortiums form simultaneously or in reaction to other formations? Would there be a domino effect? 
Is there a relation between the topology of collaboration networks and consortium composition?
 Our work indicates there may be influence of both connectivity and  diameter on the SPOA. How does the choice of a threshold (which influences the consortium size) influence SPOA? Can we propose mechanisms that further improve SPOA? The MAGNET game is a multi-round game. There are known results transforming multi-round games to single round but these involve various penalties and assumptions. Can we provide a more efficient reduction in our context? We have assumed the players to be fully rational in their decision making; it will be interesting to study such games in context of bounded rationality and also with players having limited information of their neighborhood as in a distributed network setup.

Finally, we would like to abstractly define, eventually, the class of distributed systems
formation games making it easier to understand the various trade-offs and parameters.

\section{Acknowledgements}
We would like to thank Dahlia Malkhi, Ittai Abraham and Moshe Tennenholtz for their support, and  Rann Smorodinsky, Reuven Bar Yehuda, Liron Yedidsion, Oren Ben-Zwi and the anonymous referees for their comments and discussions.

\end{document}